%% file: AAMAS_2025_CloudExploit.tex
\documentclass[sigconf]{aamas}

\usepackage{balance} 

\usepackage{appendix}
\usepackage{latexsym}
\usepackage{enumitem}
\usepackage{graphicx}
\usepackage{color}

\usepackage{times}  
\usepackage{helvet}  
\usepackage{courier}  

\usepackage{booktabs}
\usepackage{algorithm}
\usepackage{algorithmic}

\usepackage{listings}
\usepackage{xcolor}
\usepackage{soul}
\usepackage{xspace}

\newcommand{\id}{{$\mathrm{id}$}}
\newcommand{\perm}{$\mathrm{perm}$}
\newcommand{\ds}{$\mathrm{ds}$}
\newcommand{\sysname}{CloudExploit\xspace}
\newcommand{\modname}{CloudLens\xspace}
\newcommand{\us}{$\mathrm{u}$}
\newcommand{\gr}{$\mathrm{g}$}
\newcommand{\ro}{$\mathrm{r}$}
\newcommand{\<}{\langle}
\renewcommand{\>}{\rangle}

\newcommand{\mr}{\mathrm}

\newcommand{\compromised}{\mathrm{compromised}}

\newcommand{\arxiv}[1]{}

\DeclareMathOperator{\idtpl}{id\_tpl}
\DeclareMathOperator{\dstpl}{ds\_tpl}
\DeclareMathOperator{\idlongtpl}{id\_4tpl}
\DeclareMathOperator{\dslongtpl}{ds\_4tpl}
\DeclareMathOperator{\permflow}{permissionFlow}

\usepackage{newfloat}
\usepackage{listings}

\floatstyle{ruled}
\newfloat{listing}{tb}{lst}{}
\floatname{listing}{Listing}

\lstdefinelanguage{PDDL}
{
  sensitive=false,    
  morecomment=[l]{;}, 
  alsoletter={:,-},   
  morekeywords={
    define,domain,problem,not,and,or,when,forall,exists,either,
    :domain,:requirements,:types,:objects,:constants,
    :predicates,:action,:parameters,:precondition,:effect,
    :fluents,:primary-effect,:side-effect,:init,:goal,
    :strips,:adl,:equality,:typing,:conditional-effects,
    :negative-preconditions,:disjunctive-preconditions,
    :existential-preconditions,:universal-preconditions,:quantified-preconditions,
    :functions,assign,increase,decrease,scale-up,scale-down,
    :metric,minimize,maximize,
    :durative-actions,:duration-inequalities,:continuous-effects,
    :durative-action,:duration,:condition
  },
  otherkeywords = {&, |},
  keywordstyle = [2]{\color{red}},
  morekeywords=[2]{impact_attack, sensitive_data_exfiltration}
}

\title{CloudLens: Modeling and Detecting Cloud Security Vulnerabilities}

\author{Mikhail Kazdagli}
\affiliation{
  \institution{The University of Texas at Austin}
  \country{USA}
}
\email{mikhail.kazdagli@utexas.edu}

\author{Mohit Tiwari}
\affiliation{
  \institution{The University of Texas at Austin}
    \country{USA}
}
\email{mohit@symmetry-systems.com}

\author{Akshat Kumar}
\affiliation{
  \institution{Singapore Management University}
    \country{Singapore}
}
\email{akshatkumar@smu.edu.sg}

\begin{abstract}
Cloud computing services provide scalable and cost-effective solutions for data storage, processing, and collaboration. With their growing popularity, concerns about security vulnerabilities are increasing. To address this, \textit{first}, we provide a formal model, called \modname, that expresses relations between different cloud objects such as users, datastores, security roles, representing access control policies in cloud systems. \textit{Second}, as access control misconfigurations are often the primary driver for cloud attacks, we develop a planning model for detecting security vulnerabilities. Such vulnerabilities can lead to widespread attacks such as ransomware, sensitive data exfiltration among others. A planner generates attacks to identify such vulnerabilities in the cloud. \textit{Finally}, we test our approach on 14 real Amazon AWS cloud configurations of different commercial organizations. Our system can identify a broad range of security vulnerabilities, which state-of-the-art industry tools cannot detect.
\end{abstract}


\newcommand{\BibTeX}{\rm B\kern-.05em{\sc i\kern-.025em b}\kern-.08em\TeX}

\begin{document}


\pagestyle{fancy}
\fancyhead{}


\maketitle 


\section{Introduction}

Cloud computing services play a pivotal role in modernizing and optimizing businesses by providing scalable, cost-effective, and accessible solutions for data storage, processing, and collaboration. However, alongside their growing popularity, security threats in public clouds are also ever increasing~\cite{cloud42,pernetf2021}. Access control in common commercial clouds is governed by the Identity and Access Management (IAM) component, which intuitively means deciding and authorizing which users should have access to what cloud resources such as datastores, is challenging even for small and medium enterprises~\cite{aws_iam_hard}. {Amazon AWS adopts the shared responsibility model~\cite{sharedModel}, wherein the task of configuring IAM security policies is entrusted to the AWS customer.} Thus, IAM policies of an organization are configured by their IT experts who often do not have support of automated tools to analyze the full impact of their decisions on the security of the cloud configuration. As a result, several data breaches and security attacks continue to happen regularly~\cite{capitalone2022,breach1,fb_breach,linkedin_breach}.

Cloud IAM configurations (e.g., for Amazon AWS, Google GCP) are often defined using a collection of several JSON files that represent IAM policies. IAM policies encode the set of permissions that decide which user or cloud identity can access what resources (e.g., datastores). Such IAM policies create a web of identities and security permissions, which is impractical for IT admins to analyze manually. Furthermore, over time, as a result of organization changes, IAM policies move away from the principle of least privileges, and may unintentionally allow non-admin users access to sensitive data. Therefore, our goal is (a) to formally model the network of cloud identities, resources and security permissions (called \textit{\modname}) that allows reasoning about various security tasks (e.g., which users have access to what sensitive data, what types of attacks can happen such as ransomware, sensitive data exfiltration); (b) develop a tool called \textit{\sysname}, that can generate a range of security attacks in a given IAM configuration to allow vulnerability analysis and network hardening. These contributions are critical for penetration testing to harden IAM policies to common attacks such as privilege escalation, sensitive data exfiltration among others.

\paragraph{\textbf{Related work}}
There are several types of known attacks that exploit the misconfiguration of clouds such as privilege escalation attacks (a non-admin entity getting admin permissions), impact attacks (deleting data buckets), ransomware attacks (encrypting sensitive data), data exfiltration (stealing private sensitive data) among others~\cite{cloudsecthreats2023}. There has been some analysis of what kind of common IAM misconfigurations can cause such attacks~\cite{rhinosecuritylabs}. There are relatively few open source tools\footnote{Our tool \sysname will be open sourced along with anonymized instances.} that analyze a cloud configuration to detect vulnerabilities such as PMapper~\cite{nccgroup}. However, such tools do not perform a systematic search to detect complex vulnerabilities, and empirically are unable to detect a wide range of attacks our system can detect. Tools such as Policy Sentry~\cite{sentry} are preventative tools that help ease the complexity of writing secure IAM Policies. The problem of optimizing IAM policies using constraint programming has been addressed in~\cite{ijcai2022iam}. However, their method computes a new IAM policy rather than analyzes an existing one.  

Formal methods have been used to analyze the security aspect of cloud offerings by cloud providers such as Amazon. The Amazon offering Zelkova~\cite{zelkova}, an IAM policy analysis tool,  can reason whether a data bucket has public access using SMT solver. Microsoft uses theorem provers to validate global properties of their data centers~\cite{msft}. A key limitation of these tools is that they analyze the current state of the organization. In contrast, \sysname can reason about multi-step attacks that involve an attacker performing a sequence of operations to compromise the system. Another model checking method is presented in~\cite{yang2023ase}. However, this tool is specifically focused on privilege escalation (PE) attacks and can only provide a binary answer whether the current IAM configuration allows PE attacks or not. 

{
Other modeling efforts include~\cite{planning_rbac} and~\cite{policy_analysis}.
The former models only the NTFS file system, which is similar to modeling AWS S3 object storage within \sysname, however, \sysname goes beyond modeling S3 service. It accurately models permission flows that allow it to find attack paths that stem from an attacker’s ability to manipulate IAM policies (e.g. during privilege escalation or lateral movement). 
On the other hand, \cite{policy_analysis} is concerned with user-role reachability, which solves the problem of finding a path that reaches some admin role in administrative role-based access control policies (ARBAC), distinct from the role-based access control (RBAC) policies analyzed by \sysname. 
Neither of these models addresses permission flows that allow attackers to assume multiple roles, obtain group memberships, or gain elevated permissions. 
}


Attacker emulation tools such as Caldera~\cite{caldera} primarily focus on automated adversary emulation within an enterprise network context. Its primary goal is to simulate the actions of an actual attacker, making it useful for testing \textit{network defenses}. Caldera is not specifically designed to assess individual components, such as AWS IAM policies.


In the AI literature, simulated penetration testing (pentesting), for example of networked systems, is closely related~\cite{Boddy05,Obes10,SarrauteBH12,Hoffmann15}. In such a setting, a model of the underlying system is constructed and using model-based planning~\cite{FD06} attacks are generated~\cite{Hoffmann15}. Attack graphs have also been used to model complex security scenarios using a graphical network of exploits and security conditions~\cite{attackGraph,Lingyu06,Thanh18}. We will show how \modname can be used to construct such attack graphs for the IAM security model. 

\paragraph{\textbf{Contributions}}
Our key contributions are:
\begin{itemize}
    \item We develop a simple and generic representation of a cloud IAM, called \textit{\modname}, which represents security permissions, cloud identities and resources using a relation tuple-based formulation, which has been shown to be highly efficient and scalable for authorization control in large-scale clouds~\cite{zanzibar}. Key benefits of this formulation include (a) It is easily extendable to allow future cloud configuration changes; (b) it is explainable as IT admins can directly change the text file storing such tuples; (c) it is general enough to represent multiple cloud platforms such as Amazon AWS, Google GCP, Microsoft Azure among others. 
    \item We show how the problem of generating attacks (such as sensitive data exfiltration, ransomware etc) can be translated into a planning formulation~\cite{mcdermott1998pddl} with appropriate actions. Our approach is able to detect multi-step attacks. We also provide theoretical analysis of the IAM attack problem and show it is NP-Hard.
    \item Finally, we test on 14 real-world cloud instances and show that our approach can detect a range of attacks. In contrast, industry tools such as PMapper are unable to generate such a diverse range of attacks~\cite{nccgroup}.
\end{itemize}
{Our tool is uploaded as part of the supplemental files.}

\section{Cloud Identity Access Management}
\label{sec:iam}

For ease of exposition, we focus on Amazon AWS, and briefly describe relevant concepts; more details can be found here~\cite{awsDocs}. 
AWS supports both role-based access control (RBAC) and attribute-based access control (ABAC)~\cite{empirical_acl_analysis}. \modname and \sysname primarily focus on RBAC policies, which are more commonly used in practice~\cite{abac19}. In RBAC, access to cloud resources is based on a \textit{role} that usually aligns with the business logic~\cite{rbac}. ABAC policies determine access based on attributes of the subject, object, and can be converted to RBAC format~\cite{abac19}.


An \textit{IAM policy} is a set of permissions, attached to a cloud identity, that define what actions a user is allowed to execute on what AWS resources.  A resource is a cloud object with a unique identifier, such as an S3 bucket (a directory in a distributed file storage system). Each \textit{policy} consists of one or more statements and each statement describes a specific set of permissions. 

A statement in an IAM policy includes the following key elements:
\begin{itemize}
    \item \textbf{Effect:} It can be either ``Allow" or ``Deny." The ``Allow" effect grants the specified permissions, while the ``Deny" effect explicitly denies the specified permissions.
    
    \item \textbf{Action:} The specific AWS service or resource actions that the policy allows or denies. Actions can range from simple tasks like read (e.g., \texttt{s3:GetObject}) to more complex operations involving multiple resources. These actions are also called \textit{permissions}, which give the privilege of performing the associated operation. These actions are related to, but different from PDDL actions that we will define later.
    
    \item \textbf{Resource:} The AWS resources to which the actions apply. This can include specific ARNs (Amazon Resource Names).
\end{itemize}
The statement in Listing~\ref{lst:stmt} allows the ``s3:GetObject" action (which is the read permission) on the object with {Amazon Resource Name (ARN)} ``arn:aws:s3:dummyBucket'' (which is a datastore). {Every item of an IAM policy is allowed to be specified in the form of a regular expression that makes the policy able to grant multiple permissions to multiple cloud objects in a concise manner but also introduces security vulnerabilities because of making the policy over-permissive~\cite{ijcai2022iam}. 
Furthermore, each IAM policy may include an optional \textit{condition} element which acts as an additional filter, serving to further restrict the policy's scope.} 

\begin{lstlisting}[
float=t,
basicstyle=\ttfamily\small,numbers=none,
caption={IAM Policy Statement Example},
label={lst:stmt},basicstyle=\ttfamily\small]
{
  Effect: Allow,
  Action: s3:GetObject,
  Resource: arn:aws:s3:dummyBucket
}
\end{lstlisting}
An AWS policy is composed of one or more statements as shown in listing~1. Policies are associated with IAM identities, such as \textit{users}, \emph{groups}, or \emph{roles}, and they determine what actions those identities are allowed or denied to perform on AWS resources. An AWS ``identity" refers to a representation of a computing entity that can {either make requests to cloud services or control access to such services.} The key identities we address are:
\begin{itemize}
    \item \textbf{Users:} These are entities that represent individual people, applications, or services within an AWS account. Users have unique credentials (username, password) and can be assigned specific permissions using policies.
    \item \textbf{Groups:} IAM groups are collections of IAM users, roles, and other groups. By organizing identities into groups, we can manage their permissions collectively. Policies can be attached to IAM groups to grant a set of permissions to all identities within the group.
    \item \textbf{Roles:} Unlike users or groups, roles are not associated with a specific end-user or an application. Policies can be attached to roles to give them specific set of permissions. Roles can be assumed by AWS services and users, allowing them to temporarily take on the permissions assigned to the role.
\end{itemize}
In AWS, ``resources" refer to computing components such as storage resources (Amazon S3), database and networking resources~\cite{awsDocs}.

The IAM provides a centralized way to control access to different AWS resources allowing IT admins to define policies for users and other AWS identities. However, given the complexity of IAM, often misconfigurations can result which can lead to unintended side effects that result in cloud attacks. In the next section, we show one such scenario.

\subsection{Attack Example}
\label{sec:attack_example}

Consider an example where a compromised \textit{user} can execute a ransomware attack~\cite{ransomware_attack} in just a few steps. Typically, a ransomware attack involves two key actions: encrypting data and deleting the original, unencrypted data. In AWS cloud, attackers benefit from the availability of server-side encryption, which is performed within AWS, circumventing the need for conspicuous data transfers between the attacker’s machine and cloud storage. To leverage server-side encryption, an attacker requires access to the AWS Key Management Service (KMS) for encryption keys and write permissions to a relevant S3 bucket. For the sake of simplicity, this discussion assumes that the \textit{Effect} field of corresponding IAM policies is set to \textit{Allow}.


\begin{lstlisting}[
float=t,
basicstyle=\ttfamily\small,numbers=none,
caption={Ransomeware attack: policy attached to a compromised \textit{user}},
label={lst:ransomware_policy_a}]
{ Action: s3:*Object,
  Resource: s3:sensitiveDataBucket }
{ Action: sts:AssumeRole,
  Resource: iam:keyManagementRole  }
\end{lstlisting}

\begin{lstlisting}[
float=t,
basicstyle=\ttfamily\small,numbers=none,
caption={Ransomeware attack: policy attached to the role \textit{iam:keyManagementRole}},
label={lst:ransomware_policy_b}]
{ Action: kms:*Key*,
  Resource: * }
\end{lstlisting}



Policy in Listing~\ref{lst:ransomware_policy_a} grants a compromised \textit{user} ability to perform object-level operations (e.g. read/write) on the \textit{s3:sensitiveDataBucket} that contains sensitive data. However, write permission alone is not sufficient for an attacker to launch a ransomware attack. The attacker also needs to assume the \textit{iam:keyManagement} role. As permitted by the policy in Listing~\ref{lst:ransomware_policy_a} (bottom statement), this role grants extensive access to the KMS service (Listing~\ref{lst:ransomware_policy_b}). It provides the critical permission \textit{kms:CreateKey}, allowing the attacker to create a private encryption key \textit{attacker-kms-key-id}. This key, which is used for encryption but not decryption, enables the attacker to encrypt data in place using a single AWS command (Listing~\ref{lst:ransomware_aws_command}). 
\begin{lstlisting}[
float=t,
basicstyle=\ttfamily\small,numbers=none,
caption={Ransomware: in-place server-side data encryption with an attacker-controlled key \textit{attacker-kms-key-id}},
label={lst:ransomware_aws_command}
]
{ aws s3 cp s3://sensitiveData 
s3://sensitiveData --sse aws:kms 
--sse-kms-key-id attacker-kms-key-id }
\end{lstlisting}

\section{\modname: Modeling Cloud IAM}

Before attack graph and attack paths can be generated for vulnerability analysis, we need to represent different security permissions and relations among cloud identities, resources in a systematic manner. Our framework, \modname, provides such a representation. {While graph based representation of IAM has been used earlier~\cite{nccgroup}, they are limited in scope to analyzing privilege escalation attacks, and are not easy to translate into a automated planning framework}.





\subsection{Relational modeling of IAM}
\label{sec:relation}

We consider users, roles, and groups as IAM identities. For simplicity, we consider datastores as IAM resources (it is possible to consider additional resources such as compute, and network among others). At any point in time, the state of the IAM system is represented using the following 3- and 4-tuples:

\begin{itemize}
    \item \(\langle \text{{\id{1}, \perm, \id{2}}} \rangle\): This tuple implies that identity \id{1} has permission {\perm} over {\id{2}}. E.g., this permission can be \textit{assumeRole}, which implies \id{1} can assume the role \id{2} and gain all associated permissions of \id{2}. The {\id{2}} can be another identity (user, role, or group). 
    
    \item \(\langle \text{\id, \perm, \ds} \rangle\): This tuple implies that identity {\id} has permission {\perm} over datastore {\ds}. Permissions can be \text{full\_control}, \text{read}, \text{write}, \text{delete}, etc.
    
    \item \(\langle \text{\id, \id{1}, \perm, \id{2}} \rangle\): There are special types of permissions that allow an identity {\id} to add a 3-tuple  \(\tau = \langle \text{\id{1}, \perm, \id{2}} \rangle\) to the IAM state. Under certain conditions, an action can be taken (e.g., when {\id} is a compromised user) to add $\tau$ to the system state.     
    \item \(\langle \text{\id, \id{1}, \perm, \ds} \rangle\): Same interpretation as above; except that {\ds} is a datastore.
\end{itemize}
Such a simple tuple-based formulation is enough to model the majority of cloud operations in AWS, GCP~\cite{zanzibar} to analyze its security vulnerabilities. 
When we say an {\id} has permission \perm, it would refer to the tuple \(\< \text{\id, \perm, \id1} \>\) or \(\< \text{\id, \perm, \ds} \>\) ({\id1} and {\ds} would be clear from the context).

For a concrete example of a 4-tuple, consider the permission $\mathrm{iam\!:\!AttachGroupPolicy}\footnote{\url{https://docs.aws.amazon.com/IAM/latest/APIReference/API_AttachGroupPolicy.html}}$ in Listing~\ref{lst:attachgrouppolicy}. Assume identity {\id} has this permission. This permission allows {\id} to attach a policy with ARN (Amazon resource name) $\mathrm{ PID }$ to the group with ARN  $\mathrm{ GID }$. We represent this as tuple $\mathrm{\langle \text{\id}, GID, hasPolicy, PID \rangle}$.

\textbf{{Practical implementation:} }We have made a parser that mines a company's IAM policy, extracts various JSON statements analogous to listing~\ref{lst:attachgrouppolicy}, and coverts them into such a tuple format. Intuitively, such tuples become predicates in the planning formulation of the attack formulation. {Supplement (table 2) provides more information about different AWS APIs our model handles.}

\begin{lstlisting}[
float=t,
basicstyle=\ttfamily\small,numbers=none,
caption={Json for \textit{$\mathrm{iam\!:\!AttachGroupPolicy}$}},
label={lst:attachgrouppolicy}
]
{
  Effect: Allow,
  Action: iam:AttachGroupPolicy,
  Resource: <GID>,
  Condition: {
    StringEquals: {
      aws:RequestTag/policy-id: <PID>
}}}
\end{lstlisting}

\subsection{IAM Attacks as Sequential Decision Making}
\label{sec:model}

In this section, we present a minimal formal model of IAM attack problem that highlights its key aspects relating to permission flows and stripping away implementation-specific details.

\paragraph{State definition}

At any instant $t$, the IAM state $s_t$ is a \textit{set} of tuples as noted in section~\ref{sec:relation}. In addition, we also have predicates $\compromised(\text{\id})$ which are true for identities that are compromised (i.e., the attacker controls them).

Let the current state be $s$. We define the following action types:
\begin{description}
   \item[$\boldsymbol{\permflow(\text{\id{2}}, \text{\id{1}})}$] This action is applicable when permissions can flow from id2 to id1 given the current state $s$. 
   
   \textbf{Action precondition: }Let the precondition for this action be ${\mathrm{isFlowActive}(\text{\id{2}}, \text{\id{1}})}\subseteq s$. It can be defined based on the interpretation of different permissions in the specific cloud type (e.g., AWS, GCP). Also note that this action can be instantiated for all possible identities {\id{1}}, {\text{\id{2}}}. 

   Intuitively, ${\permflow(j,i)}$ is equivalent to identity $i$ assuming the role of identity $j$, which leads to permission flow from $j$ to $i$. In a particular attack plan, an identity $i$ can only assume one role. Thus, if action ${\permflow(j,i)}$ is taken, then no other action ${\permflow(k,i)}$ (and $k\neq j$) can be taken. This can be easily ensured by using a predicate $\mathrm{inFlow(i)}$ which is false initially and becomes true after an action ${\permflow(j,i)}$ is taken (details omitted).
   
   \textbf{State Transition: }The effect is that all the permissions tuples $\text{\id{2}}$ has flow to $\text{\id{1}}$. Given state $s$, the transition function $s'=T(s, \permflow(\text{\id{2}}, \text{\id{1}}))$ is given as:
\begin{align}
    \hspace{-8pt}s' \leftarrow s &\cup \{\<\text{\id{1}}, \text{\perm}, \text{\id}\>\; \forall  \<\text{\id{2}}, \text{\perm}, \text{\id}\>\in s\} \nonumber \\
    &\cup \{\<\text{\id{1}}, \text{\id{3}}, \text{\perm}, \text{\id{4}}\> \; \forall \<\text{\id{2}}, \text{\id{3}}, \text{\perm}, \text{\id{4}}\>\in s\}
\end{align}
Datastore specific permissions also flow from {\id{2}} to {\id{1}} analogously (omitted for clarity).

\item[$\boldsymbol{\mathrm{addTuples}(\text{\id})}$] This action acts on the 4-tuples $\langle {\text{\id}, \text{\id{1}}, \text{\perm}, \text{\id{2}}} \rangle$, and adds the tuple $\langle \text{\id{1}}, \text{\perm}, \text{\id{2}} \rangle$ to the state. This action lets an attacker directly modify the IAM state.

    \textbf{Action precondition: }The {\id} must be the compromised. That is, predicate $\mathrm{compromised}(\text{\id})\in s$ must be hold. 

    \textbf{State Transition: } The transition function is given as:
        \begin{align}
            s' \leftarrow s &\cup \{\<\text{\id{1}, \perm, \id{2}}\> \; \forall \<\text{\id, \id{1}, \perm, \id{2}}\>\in s\}  
        \end{align}
    Datastore-specific permissions are also added analogously (omitted for clarity).
        
\item[enableAttack({\id}, attack\_type)] This action lets {\id} execute the attack\_type (there can be multiple attack types possible such as sensitive\_data\_exfiltration, ransomware etc).

\textbf{Action precondition: } The precondition for this action depends on the attack type to be executed and is typically given by the presence of tuples of certain types in the current state $\mathrm{precondition(\text{\id}, attack\_type)}\in s$. These preconditions can be defined for different attack types. Also, we have $\compromised(\text{\id})\in s$.

\textbf{State Transition: }This a terminal action. Once executed, the attack is assumed to be successful.
\end{description}

\paragraph{Goal} The cost of each action is one. The goal is to find the least cost valid sequence $s_0, a_0, \ldots, s_n, a_n$ where the last action $a_n$ is $\mathrm{enableAttack(\cdot)}$. Each action $a_t$'s precondition must also be satisfied in the state $s_t$.


Finding a low cost plan is useful as practically we observed that often redundant, benign actions (such as reading a public datastore) can be added to make the plan arbitrarily large. Having such redundant actions makes attack plans less useful and imprecise for end users. Based on this flow based formulation, we also show in supplement that the IAM attack path problem is NP-Hard.

We incorporate a list of some known attacks in Table~\ref{tab:attacks} (\emph{bucket} refers to an object storage container like a directory); however, our planning framework is generic and can be extended to future attack types as we model actions that allow flow of permissions among entities, which is a key source of multi-step complex attacks.

\begin{table}[t]
    \small 
    \centering
    \small
    \begin{tabular}{|p{0.4\linewidth}| p{0.5\linewidth}|}
         \hline
        \textbf{Attack type} & \textbf{Interpretation} \\ \hline
        Sensitive data exfiltration & Move sensitive data to a public bucket\\ \hline 
        Impact & Delete bucket, user, role, group, or policy\\ \hline 
        Persistence & Create a new IAM user \\ \hline 
        Lateral movement  & Set or change password/access key for a user\\ \hline 
        Privilege escalation  & Get elevated privileges for a user\\ \hline 
        Ransomware & Encrypt sensitive data and delete the original data\\ \hline
    \end{tabular}
    \caption{Attack Types}
    \label{tab:attacks}
\end{table}

\subsection{Attack graph formulation}
\label{sec:attackgraph}

We can construct an \textit{attack graph}, analogous to the planning graph used by the graph plan algorithm for classical planning problems~\cite{blum97}. Intuitively, the plan graph is a layered graph consisting of alternating levels of:
\begin{itemize}
\item State level -- Represents tuples (as in section~\ref{sec:relation}) that can be true at that level. At the initial level, only those tuples that are true are listed. These tuples are analogous to security conditions in an attack graph.
\item Action level -- Represents the actions that can be executed at that level based on the tuples (i.e., action preconditions) from the previous level. An action has a link from each of its precondition, and a link to its effect (as noted in previous section).
\item Effect propagation -- Each action contributes its effects (the tuples it adds or deletes) to the next state level.
\end{itemize}
There are additional concepts such as mutually exclusive (\textit{mutex}) actions, tuples, and iteratively deepening the planning graph which are detailed in~\cite{blum97}. This graph is analogous to an attack graph where finding a path (a sequence of actions) until action \textit{enableAttack} can be executed solves the attack generation problem. 


\section{\sysname: Automated Planning for Attack Generation}
\label{sec:permflow}

{As noted in previous section, a key challenge is to identify a sequence of actions (e.g., cloud API calls) that can achieve a malicious objective (e.g., launching an impact attack). This task---generating an action sequence to reach a goal---naturally aligns with AI planning. The ability to generate such action sequences is crucial for verifying the existence of security vulnerabilities. While model checking is used in previous cloud security works~\cite{yang2023ase}, formal methods typically provide a binary answer regarding the presence of a safety property, and cannot easily handle action costs. Therefore, we formulate the planning problem in section~\ref{sec:model} using the planning domain definition language (PDDL)~\cite{mcdermott1998pddl}, which has been used earlier also for network security and penetration testing~\cite{Boddy05,Obes10,SarrauteBH12,Hoffmann15}}.

{PDDL is a standardized language used to describe planning problems and domains in automated planning systems~\cite{mcdermott1998pddl}. It provides a formal way to define the states, actions, and goals for planning tasks. It separates the description of a planning problem into two main parts: the domain (which includes predicates, actions, and their effects) and the problem (which specifies the initial state and goal conditions). For more details on PDDL syntax, we refer to~\cite{pddl_wiki}.}

Based on the simplified model in the previous section, we now describe specific implementation of different actions for Amazon AWS in PDDL. {We first parse an organization's cloud setup to construct the relational model as in Section~\ref{sec:relation}}. We next show how the planner (or the attacker) uses such relations starting from a compromised user to make progress towards accomplishing an attack (as noted in Table~\ref{tab:attacks}). The key to gain permissions to enable an attack is via permission flow between identities as described next.

Assume we have three identities in the cloud---user \us{1}, group \gr{1} and role \ro{1}; and one datastore {\ds{1}}. Assume the tuples that describe the current state are:
\begin{align*}
    &\<\text{\us{1}}, \textrm{belongsTo}, \text{\gr{1}}\> \\
    & \< \text{\gr{1}}, \textrm{any\_user}, \textrm{assumeRole}, \text{\ro{1}} \> \\
    & \< \textrm{\ro{1}}, \textrm{deleteBucket}, \text{\ds{1}} \>
\end{align*}
For the impact attack to succeed, {\us{1}} needs to delete {\ds{1}}. However, in the current state, {\us1} does not have the target permission $\< \textrm{\us1, deleteBucket, \ds1} \>$. The following sequence of operations takes place to enable this attack:

\begin{itemize}
  \item Since, {\us1} belongs to group {\gr1}, it assumes all the permissions of {\gr1}. After this flow, the tuple $\< \textrm{\us{1}, any\_user, assumeRole, \ro{1}}\>$ is activated.
  \item We use the keyword ``any\_user" to denote that any user in the system can replace it. Therefore, \us{1} replaces any\_user with itself and activates the tuple $\< \textrm{\us1, assumeRole, \ro{1}}\>$.
  \item Tuple $\< \textrm{\us1, assumeRole, \ro{1}}\>$ allows {\us1} to assume the role {\ro{1}} and thus gain all the permissions role {\ro{1}} has. In this way, {\us1} gains target permission $\< \textrm{\us1, deleteBucket, \ds1} \>$.
\end{itemize}

From this example, one can see how even for a system with a small number of tuples, permission flow can quickly become challenging. Therefore, we need an automated planning method to systematically search through all permission flow paths and check if an attack is possible. This is achieved by the PDDL modeling of the problem. We categorize actions into four classes as defined next.

\paragraph{\textbf{Permission flow actions}} We first focus on PDDL actions that enable different kinds of permission flow. In what follows, we shall introduce relevant PDDL predicates alongside the actions that require them. The predicates below indicate the corresponding relation tuples introduced in section~\ref{sec:relation}.
\begin{itemize}
    \item $\idtpl (\text{?\id1, ?\perm, ?\id2})$ 
    \item $\dstpl(\text{?\id1, ?\perm, ?\ds})$
    \item $\idlongtpl(\text{?\id, ?\id1, ?\perm, ?\id2})$
    \item $\dslongtpl(\text{?\id, ?\id1, ?\perm, ?\ds})$
\end{itemize}


\begin{lstlisting}[
  float=t,
  caption={\small Permission flow of 3-tuple relations from \id2 to \id1},
  label={lst:permFlow},
  language=PDDL,numbers=none,basicstyle=\ttfamily\small]
(:action permissionFlow_id_3tpl
  :parameters (?id2 ?perm ?id ?id1)
  :precondition 
  (
    isFlowActive(?id2, ?id1)&
    id_tpl(?id2, ?perm, ?id)&
    not id_tpl(?id1, ?perm, ?id)
  )
  :effect id_tpl(?id1, ?perm, ?id)
)
\end{lstlisting}
For exposition ease, we introduce the following logical condition that checks whether permissions can flow from \id2 to \id1:
\begin{align}
   & \mr{isFlowActive} \text{(?\id2, ?\id1)} := &\nonumber \\
   & \idtpl\text{(?\id1, $\mr{assumeRole}$, ?\id2)} \; &\vee \nonumber \\
   &  \idtpl\text{(?\id1, $\mr{belongsTo}$, ?\id2)} \; &\vee \nonumber \\
   &   \idtpl\text{(?\id1, $\mr{hasPolicy}$, ?\id2 )} &
\end{align}
This logical condition ensures if \id1 can either assume the role \id2 (if \id2 is a role), or \id1 is part of \id2 (if \id2 is a group), or in case \id2 is a policy (a policy in AWS can also be represented as an identity type), then \id1 has the policy \id2.

The interpretation of the action \texttt{permissionFlow\_id\_3tpl} in Listing~\ref{lst:permFlow} (using $\mr{isFlowActive}$) is: 
\begin{itemize}
    \item Assume that \id2 has permission $\idtpl\text{(\id2, \perm, \id)}$.
    \item Assume $\mr{isFlowActive}$ \text{(\id2, \id1)} holds.
    \item In such a case, the permission that \id2 has ($\idtpl\text{(\id2, \perm, \id)}$) can flow to \id1. This implies the tuple $\idtpl\text{(\id1, \perm, \id)}$ will become activated as an effect of this action.
\end{itemize}
We can define an analogous action \texttt{permissionFlow\_ds\_3tpl} for permissions over datastores (or resources). Permissions flow for 4-tuple relations are defined analogously in the supplement. It is possible to have a permission flow action that flows  all the permissions \id2 has to \id1 at once. In our testing, while simpler, this action used \textit{forall} effects, which was challenging for the planning solver. Therefore, we use the current design of permission flow actions. 



\paragraph{\textbf{Actions for handling regular expressions}}For succinct and efficient representation in PDDL (w.r.t.$\!$ the number of tuples), we use the keyword ``full\_control" to match to \textit{any} permission in the system, and ``any\_user" to match to any user (or analogously any role/group) in the cloud environment. Such permissions are useful for defining admin users or admin policy, or succinctly defining a set of permissions an identity can have. This representation also helps the planner as the planner only needs to instantiate the required tuple for an attack to match these keywords (and not explicitly create all the tuples). It is shown in supplement.

\begin{lstlisting}[
  float=t,
  caption={\small Action for adding a 3-tuple to the cloud environment},
  label={lst:activate_id_4tpl},
  language=PDDL,numbers=none,basicstyle=\ttfamily\small]
(:action add_id_3tpl
    :parameters (?id1 ?id2 ?perm  ?id3 )
    :precondition (
        not id_tpl(?id2 ?perm  ?id3) &
        compromised(?id1) &
        id_4tpl(?id1 ?id2 ?perm  ?id3 )
        )
    :effect id_tpl(?id2 ?perm  ?id3)
)
\end{lstlisting}




\paragraph{\textbf{Actions for adding additional tuples}} An attacker in a cloud setting can also add additional relations to the environment to help itself achieve target attacks. This can be done by using 4-tuple relations as described in Section~\ref{sec:relation}. Action in Listing~\ref{lst:activate_id_4tpl} checks if \id1 trying to add a tuple is a compromised id, has the required permission $(\text{\id1, \id2, \perm, \id3})$, and tuple to be added is not already present, then the effect makes the predicate $\mr{id\_tpl}(\text{\id2,\perm, \id3})$ true. That is, now \id2 has permission {\perm} on \id3, which was not present earlier in the system. Analogously, we can define an action \texttt{add\_ds\_3tpl} to add additional permissions over datastores.

\paragraph{\textbf{Actions to enable attacks}}Previous class of actions described how permissions flow among different identities, how to expand regular expressions and add additional relations to the cloud environment. Actions in the current class check the set of permissions the attacker has to enable any of the attacks noted in Table~\ref{tab:attacks}. An example action for sensitive data exfiltration attack in shown in supplement. Other such actions are in our PDDL source code repository. 




\paragraph{\textbf{Initial state}} The initial state in the PDDL formulation consists of all the 3- and 4-tuple relations mined from an organization's AWS IAM. It also contains information about which datastores are public and/or have sensitive data. We also have predicates for currently non-existent datastores, which can be created by the attacker for a data exfiltration attack; same holds for non-existing user identities for the persistence attack. There are also predicates denoting which identities are users, roles, groups, datastores, or policies. 

We also need to define which user is compromised. There can be multiple ways to do this. We can run the planner by setting each user as compromised one by one. Another method, which is faster, when we are looking to generate any attack plan for a particular type of attack is to let the planner choose which user to make compromised. This action is \texttt{selectCompromisedUser} shown in supplement.


\paragraph{\textbf{Goal state}} The goal state contains which attack type we would like to generate an attack for (that is, a  predicate for each entry in Table~\ref{tab:attacks}). 


\section{Results}
Our evaluation methodology is twofold, addressing both the verification of modeling accuracy and the identification of actual attacks. For verification, we employ the IAM Vulnerable benchmark~\cite{iam_vulnerable} developed by Bishop Fox~\cite{bishop_fox}, a penetration testing firm. This benchmark comprises 31 attack scenarios associated with various AWS cloud services, particularly those vulnerable to misconfigurations potentially leading to privilege escalation attacks. 
The accuracy of the model is demonstrated through the application of \sysname across 14 distinct cloud configurations belonging to various commercial organizations. Details about the experimental setup are in supplementary.
Anonymized datasets used in our evaluation together with \sysname's implementation will be publicly released.


\begin{figure}[tbp]
\centering
\includegraphics[width=0.5\textwidth]{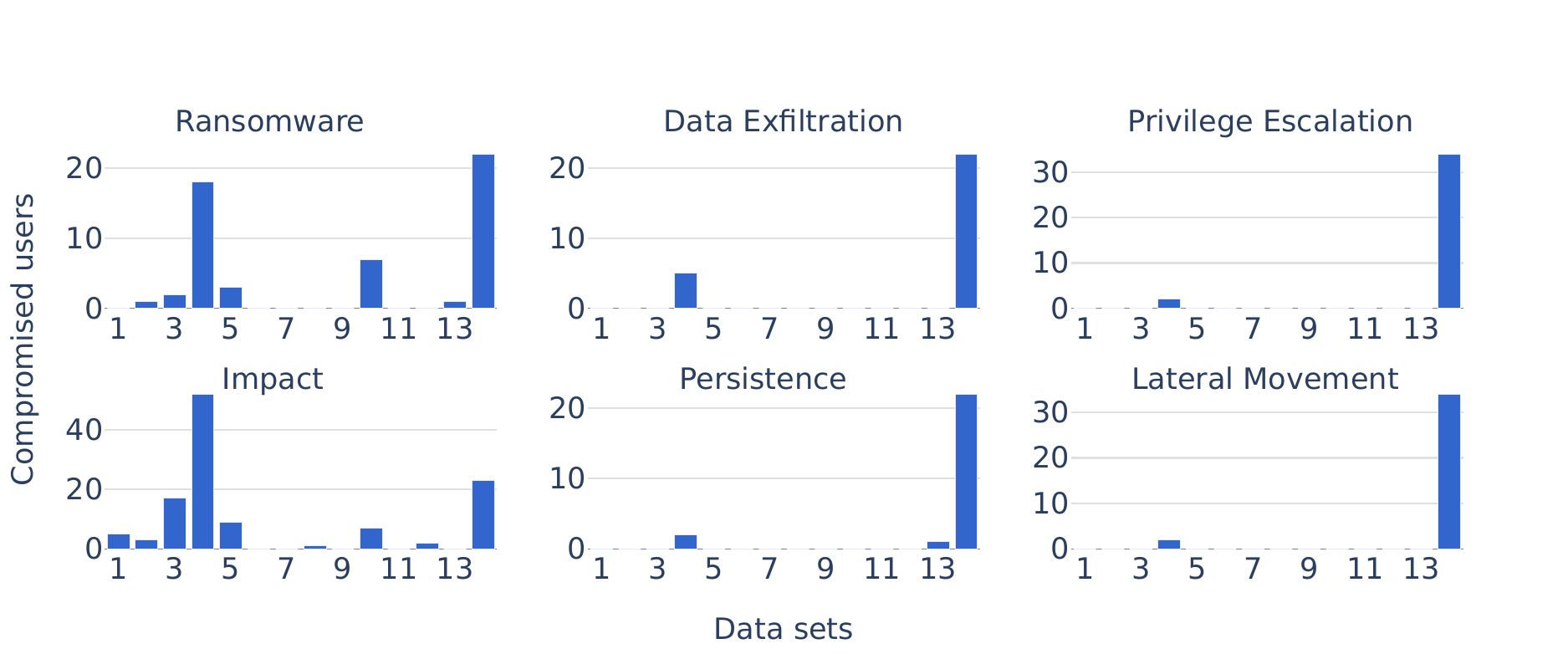}\par
\caption{Real-world datasets: The number of unique compromised users (after exclusion of admin users\arxiv{1cm}).}
\label{fig:affected_users}
\end{figure}

\subsection{IAM Vulnerable Benchmark}
We conducted a thorough evaluation of our model’s accuracy against IAM Vulnerable benchmark~\cite{iam_vulnerable}, specifically focusing on the attack scenarios relevant to our modeling efforts. Remarkably, \sysname successfully detected all 19 attacks (out of 31) that are within the scope of PDDL modeling. Our code base provides PDDL problem data files for this benchmark.

In terms of IAM policy modeling, \sysname demonstrates comprehensive support for the most frequently used services, such as IAM, EC2, S3, Lambda, STS, and SSM. However, we deliberately chose to exclude less common services like SageMaker, CloudFormation, CodeBuild, and DataPipeline from our modeling process to keep PDDL problem files compact for solver scalability. This exclusion accounts for the 12 undetected test cases in the benchmark. In summary, our evaluation highlights \sysname{}'s effectiveness in modeling and detecting vulnerabilities within the scope of widely used AWS services.

\subsection{Real-world Datasets}
\sysname{} is employed to analyze 14 commercial cloud configurations. Details of these cloud configurations are in the supplement. 
The data includes hundreds to thousands of identities and datastores, and several thousand permissions. To our knowledge, this dataset is one of the most extensive in cybersecurity for modeling cloud permission vulnerabilities. 


To enable Fast Downward~\cite{FD06} solver to process such large datasets, we partition them based on AWS accounts that serve as effective boundaries for both administration and billing. In some cases account partitioning is not sufficient because the solver still runs out of memory. To handle those situations we partition an account into non-overlapping groups of users and include other types of identities and datastores that are reachable from the source users via permission edges using breadth-first search. We intentionally exclude admin users from our analysis, as compromising such users is likely to have a widespread impact on the entire system, attributable to their extensive array of potentially high-risk permissions.



\begin{figure}[t]
\centering
\includegraphics[scale=0.3]{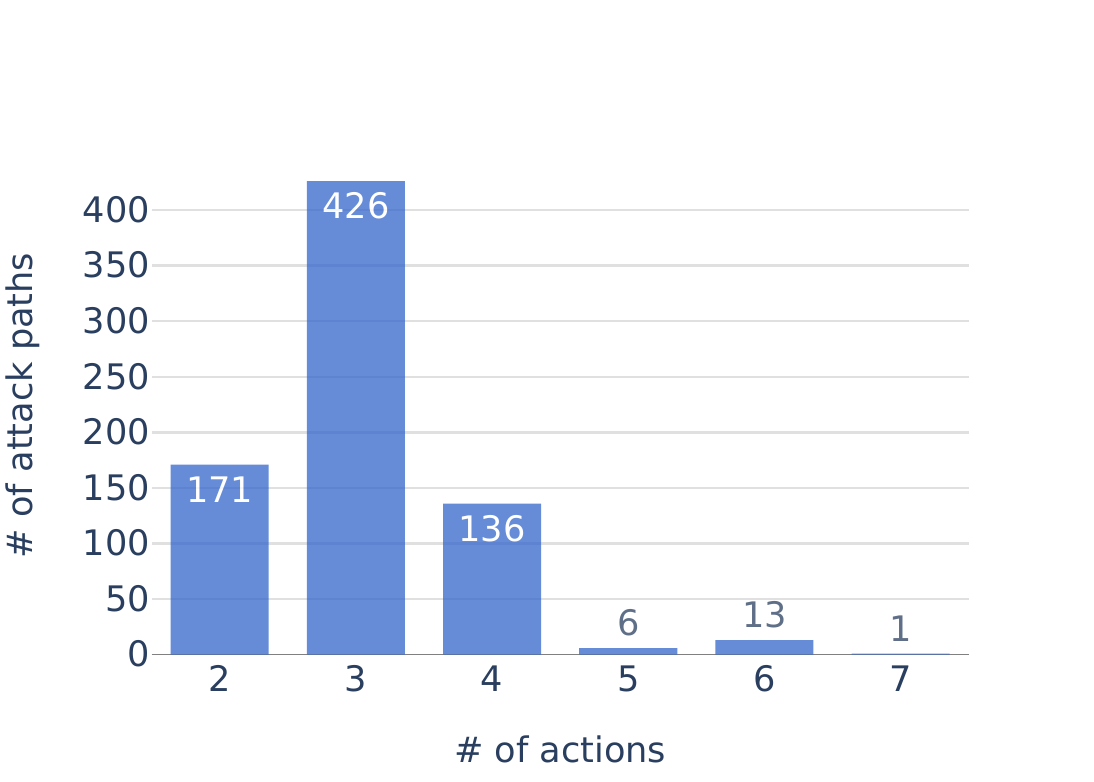}\par
\caption{Distribution of attack path lengths measured in terms of the number of actions\arxiv{1cm})}
\label{fig:attack_path_distribution}
\end{figure}

\paragraph{Attack path diversity} Our evaluation methodology is designed to estimate the diversity of attack paths generated by \sysname. It is important to note that a mere syntactic analysis of these paths is insufficient, as their diversity could be artificially increased by the inclusion of irrelevant actions that do not alter their semantic meaning. For instance, a path that incorporates a role assumption action that is irrelevant to the attack may appear different from one without such a redundant element, yet both paths are semantically identical. Our approach to measuring attack diversity draws on principles from the cybersecurity domain. We specifically gauge the number of compromised users for each attack-data set combination. While there may exist other diverse paths, the critical factor for our analysis is the extent to which users are vulnerable to attacks.

Figure~\ref{fig:affected_users} presents the number of compromised users for each attack, data set combination. The  y-axis indicates the count of compromised users, while the data sets are arrayed along the x-axis. 
\sysname identified 507 non-admin users across the datasets who could potentially be exploited by attackers to initiate one or more attacks if compromised.

A noticeable trend is the vulnerability of almost all real data sets to ransomware and impact attacks, such as encrypting and deleting sensitive data or removing critical cloud resources: identities, access keys essential for programmatic access to cloud resources, and IAM policies. The latter action usually leads to the removal of access-control constraints, potentially resulting in privilege escalation.

At least one user in nearly every environment can execute a ransomware attack. This vulnerability stems from flaws in the IAM model. To conduct a ransomware attack, as detailed in Section~\ref{sec:attack_example}, a user requires just two permissions: s3:PutObject and kms:CreateKey. The former is a common `write' permission, while the latter is often inadvertently assigned through blanket KMS access, neglecting the exclusion of hazardous permissions. Additionally, \sysname evaluates two critical S3 bucket configuration parameters: versioning and multi-factor authentication (MFA). The activation of bucket versioning enables the restoration of any version of objects stored in an S3 bucket. Meanwhile, MFA protection mandates extra authentication for object deletion in an S3 bucket, enhancing security against unauthorized actions. In our experiments, we found that both of those configuration settings are disabled sometimes, which is a prerequisite for a successful ransomware attack.

Contrary to our initial expectations, our analysis identified that three organizations are prone to privilege escalation attacks, which are among the most severe security vulnerabilities. 
Surprisingly, in the data set \#5, numerous users have the capability to carry out data exfiltration attacks. A deeper investigation disclosed that the organization possesses an S3 datastore filled with synthetic sensitive data accessible to almost all users for software development purposes.


\paragraph{Distribution of attack path lengths}
In our analysis, we focus on the lengths of the generated attack paths, as illustrated in Figure~\ref{fig:attack_path_distribution}. Predominantly, these paths are short, with lengths of 2 or 3 actions, indicating that compromised users often have immediate access to the necessary permissions for launching attacks. Conversely, paths extending to 4 or more steps represent a distinct pattern, typically involving the propagation of permissions via PDDL permission flow actions (Section~\ref{sec:permflow}). From a security perspective, this indicates a requirement for attackers to augment their permissions, either through assuming roles or by leveraging inherited permissions from their security group memberships. 

\begin{figure}[t]
\centering
\includegraphics[scale=0.35]{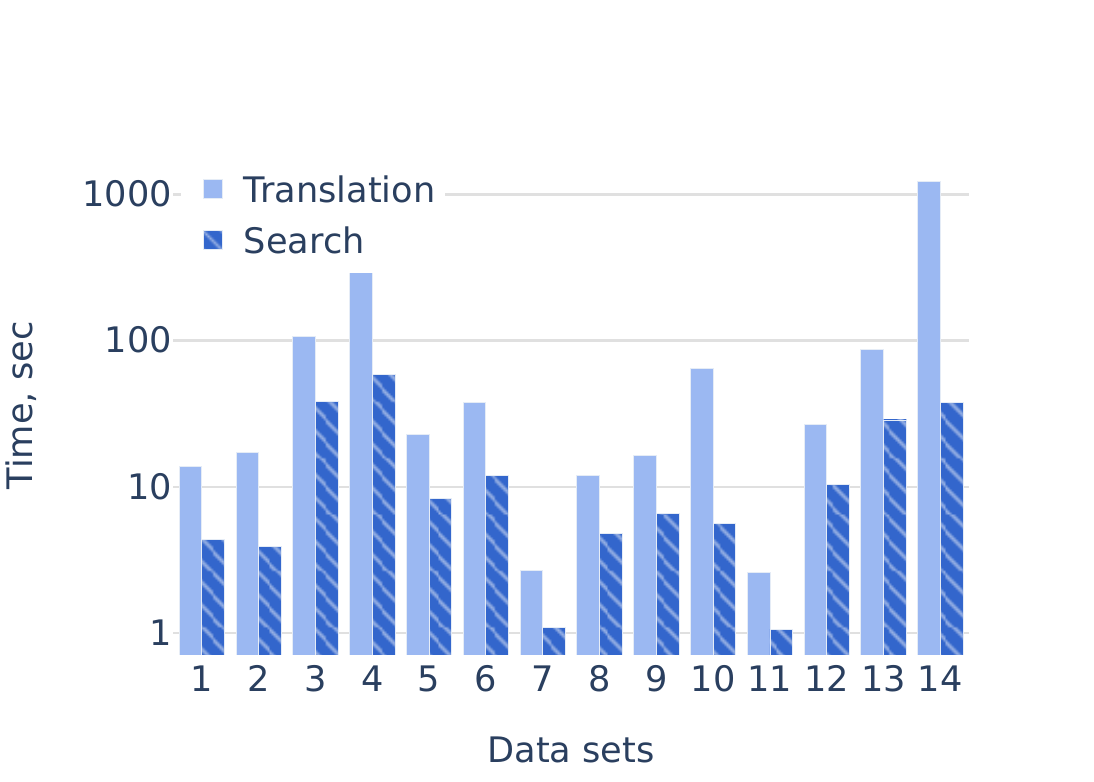}\par
\caption{Execution time of the translation and search phases of the Fast Downward planner (log scale)\arxiv{1cm})}
\label{fig:execution_time}
\end{figure}

\paragraph{Execution time}

We mentioned earlier the Fast Downward solver's scalability issues with real-world datasets. Our analysis, detailed in Figure~\ref{fig:execution_time}, examines the execution times of the translation and search phases separately, with results plotted on a logarithmic scale on the vertical axis. Significantly, the translation phase dominates in execution time and is more prone to memory exhaustion compared to the search phase.
Furthermore, we conducted two experiments using CPDDL planner~\cite{cpddl}, employing both its translation algorithm and lifted planning approach. However, neither experiment yielded performance improvements over Fast Downward planner.
Improving scalability would result in even more attacks being found.

\paragraph{Real-world Attack Examples}
As previously mentioned, we conducted a manual analysis of generated attack paths to ensure they were free of modeling errors. In this section, we present a selection of these attack paths for illustration. For clarity, we focus on the shorter paths. We exclude the discussion of ransomware attacks, as their details have been covered in Section~\ref{sec:attack_example}. {The supplement shows a more complex, longer plan.}

\paragraph{Impact attack} Listing~\ref{lst:impact_attack} demonstrates a three-step impact attack. 
The attack initiates by designating \textit{user\_181} as a compromised entity. It then enables the \textit{deleteBucket} permission. 
This permission is available to \textit{user\_181} due to their comprehensive (\textit{full\_control}) access rights, thus authorizing them to employ the \textit{deleteBucket} action on \textit{data\_store\_71}. This particular S3 bucket, identified as containing sensitive data through a prior data scan, becomes the target. The attack reaches its conclusion when \textit{user\_181} executes the \textit{deleteBucket} action, effectively compromising the data.

\begin{lstlisting}[
  float=b,
  caption={\small Impact attack},
  label={lst:impact_attack},
  language=PDDL,numbers=none,basicstyle=\ttfamily\small]
(:action selectCompromisedUser
  :parameters (user_181))
(:action activate_ds_3tpl
  :parameters (user_181 
  deleteBucket data_store_71))
(:action deleteBucket
  :parameters (user_181 data_store_71))
\end{lstlisting}

\paragraph{Sensitive data exfiltration}
We discovered that in one of the commercial environments (referred to as dataset \#5), there exists an IAM configuration that permits \textit{user\_0} access to \textit{data\_store\_0}, a repository of sensitive data, as well as to a public \textit{data\_store\_138} (Listing~\ref{lst:sensitive_data_attack}). At first glance, this might appear as a minor issue in IAM design. However, this configuration could potentially lead to a significant data breach. The absence of protections against legitimate data transfers between S3 buckets means attackers could transfer large volumes of sensitive data to \textit{\textit{data\_store\_138}} rapidly. AWS S3, known for its high data throughput capabilities, could facilitate the swift transfer of gigabytes of sensitive information out of public \textit{data\_store\_138}.


\begin{lstlisting}[
  float=t,
  caption={\small Sensitive data exfiltration attack},
  label={lst:sensitive_data_attack},
  language=PDDL,numbers=none,basicstyle=\ttfamily\small]
(:action selectCompromisedUser
  :parameters (user_0))
(:action copyobject
  :parameters (user_0 data_store_0 
  data_store_138))
\end{lstlisting}

\subsection{Comparison against PMapper} 

\begin{figure}[htb]
\centering
\includegraphics[scale=0.4]{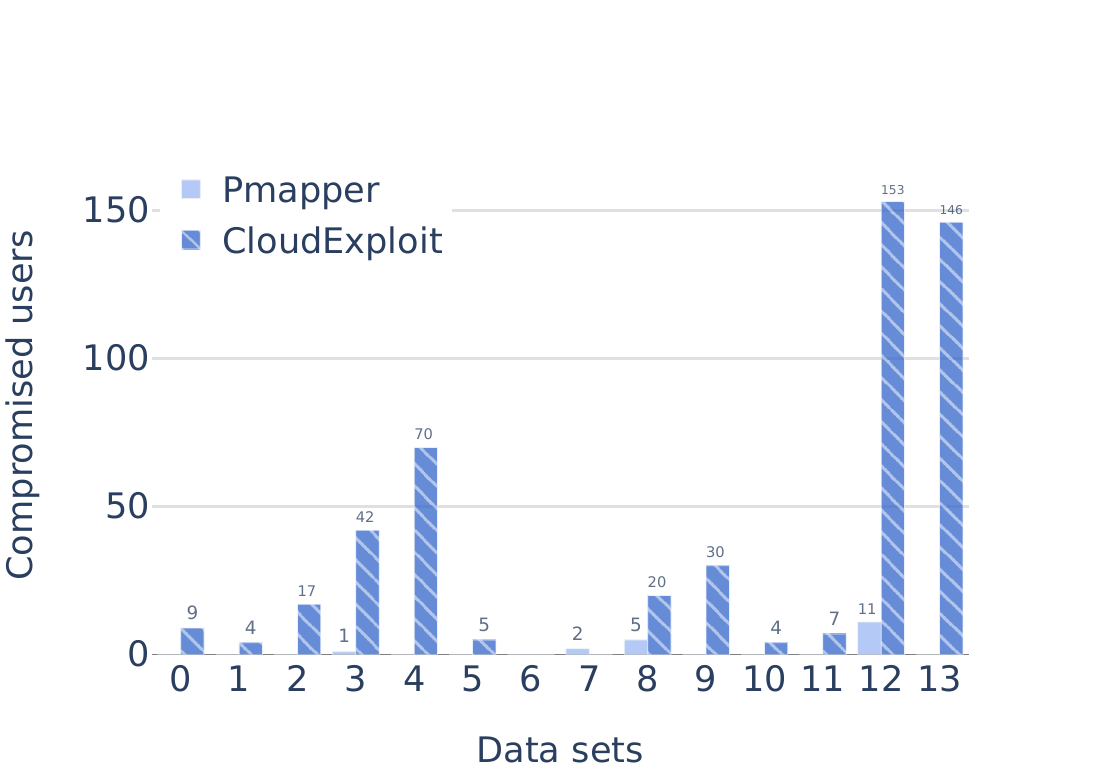}\par
\caption{Comparison of PMapper and \sysname{} \arxiv{2cm})}
\label{fig:cloud_exploit_vs_pmapper_results}
\end{figure}

We compare PMapper~\cite{nccgroup}, a popular publicly available tool tackling a similar problem as \sysname. 
{
According to the study~\cite{pmapper_assessment}, PMapper offers the most comprehensive detection capabilities and demonstrates significantly higher accuracy compared to other publicly available tools. 
It supports the widest range of unique privilege escalation paths and correctly handles the greatest number of outlined capabilities, resulting in fewer false positives and false negatives.
}

{
In comparison to PMapper, which is limited to identifying only \textit{privilege escalation attack} paths, \sysname has a much broader scope - it detects 6 different attack types.} Unlike \sysname's formal methodology for modeling IAM permissions, PMapper relies on scanning IAM policies and identifying dangerous permissions via ad-hoc definitions, making it incapable of detecting attacks involving sensitive data, such as ransomware or data exfiltration, due to its lack of datastore visibility. \sysname detected 507 users vulnerable to one or more of six modeled attack types, whereas shown in Figure~\ref{fig:cloud_exploit_vs_pmapper_results}, Pmapper detects significantly fewer compromised users. This shows the strength of our method over such existing approaches.


\section{Future work} 
Diverse plans can be useful to identify the weakest precondition(s) that enable such attacks, and remove them. 
We tested the diverse planner~\cite{diverse_planner}, but found that generated plans were not useful as the planner added benign actions (e.g., list files in a directory), while keeping the same core attack path. It can be interesting to develop methods for minimal diverse plans. 

A fully automated approach to policy repair necessitates additional information, such as the critical services required by different users, to ensure that essential accesses are not inadvertently revoked during the repair operation. We envision \sysname as a security co-pilot that IT administrators can utilize to assess the attack surface and proactively address potential policy misconfigurations.


\section{Conclusion}
{
Our research not only identifies key vulnerabilities in contemporary cloud infrastructures but also offers a viable and efficient solution for mitigating these risks, thus significantly improving the security and dependability of cloud computing services. \sysname, grounded in a robust formal framework, addresses existing deficiencies in cloud security. Its proficiency in detecting sophisticated cloud attacks has been thoroughly validated across multiple commercial cloud configurations.
}



\begin{acks}
If you wish to include any acknowledgments in your paper (e.g., to 
people or funding agencies), please do so using the `\texttt{acks}' 
environment. Note that the text of your acknowledgments will be omitted
if you compile your document with the `\texttt{anonymous}' option.
\end{acks}


\bibliographystyle{ACM-Reference-Format} 
\bibliography{aaai24}

\newpage 
\begin{appendices}
   \input{appendix}
\end{appendices}

\end{document}

%% file: appendix.tex




\section{Long Attack Path} 

\begin{lstlisting}[
  float=b,
  caption={\small An example of a 6-step attack. The compromised \textit{user\_9} attaches \textit{adminPolicy} that allows to execute all 6 attacks},
  label={lst:5_step_admin_attack},
  language=PDDL,numbers=none,basicstyle=\ttfamily\small]  
(:action selectCompromisedUser
  :parameters (user_9))
(:action permissionFlow_id_4tpl
  :parameters (role_10, role_13, 
  role_10, hasPolicy, adminPolicy))
(:action permissionFlow_id_4tpl
  :parameters (user_9, role_10, 
  role_10, hasPolicy, adminPolicy))
(:action add_id_3tpl
  :parameters (user_9, role_10, 
  hasPolicy, adminPolicy))
(:action permissionFlow_id_3tpl
  :parameters (role_10, hasPolicy, 
  adminPolicy, user_9))
(:action reachAdminPolicy
  :parameters (user_9))
\end{lstlisting}


To illustrate the effectiveness of our planning approach, we demonstrate a six-step attack (Listing~\ref{lst:5_step_admin_attack}) found in dataset \#14.
The attack begins with \textit{user\_9} who possesses permission to assume role \textit{role\_10}, which, in turn, can be used to assume \textit{role\_13}. 
Notably, \textit{role\_13} is endowed with dangerous permission \textit{iam:AttachRolePolicy} that enables the execution of a broad spectrum of cloud actions.
We model the \textit{iam:AttachRolePolicy} permission as one that allows the attachment of \textit{adminPolicy}, which comes with unrestricted permissions. 
Consequently, \textit{user\_9} can achieve unrestricted permissions through a sequence of role assumptions.

The attack unfolds in the following manner:
\begin{itemize}
    \item  \textit{selectCompromisedUser}: The planner marks \textit{user\_9} as compromised.
    \item  \textit{permissionFlow\_id\_4tpl}: It establishes a permission flow from \textit{role\_13} to \textit{role\_10} using \textit{sts:AssumeRole}, enabling the tuple \(\langle \textit{role\_10, role\_10, hasPolicy, adminPolicy} \rangle\).    
    \item  \textit{permissionFlow\_id\_4tpl}: It establishes a permission flow from \textit{role\_10} to \textit{user\_9} using \textit{sts:AssumeRole}, enabling the tuple \(\langle \textit{user\_9, role\_10, hasPolicy, adminPolicy} \rangle\).
    \item  \textit{add\_id\_3tpl}: A 3-tuple  \(\langle \textit{role\_10, hasPolicy, adminPolicy} \rangle\) is added to the environment by the planner.
    \item  \textit{permissionFlow\_id\_3tpl}: The planner applies permission flow from \textit{role\_10} to the compromised \textit{user\_9} by enabling the tuple \textit{user\_9, hasPolicy, adminPolicy}.
    \item  \textit{reachAdminPolicy}: \textit{user\_9} gains \textit{adminPolicy}, providing broad permissions and enabling all attack types modeled by \sysname.
\end{itemize}

\section{Experimental Setup}
{\sysname} has been implemented in Python. The system retrieves cloud IAM configurations and conducts comprehensive data preprocessing. The preprocessing step involves expanding regular expressions and narrowing the scope of analysis to align with the security constraints inherent in the AWS security model. We then solve the formulated PDDL problem instances using the Fast Downward
solver, running in AWS cloud on r6a.2xlarge instances equipped with 8 vCPU cores and 64GB of RAM. Furthermore, we also did manual verification of majority of attacks (specially the long horizon attacks) to ensure the absence of any modeling errors.

\begin{table}[t]
\small 
\begin{center}
\begin{tabular*}{0.46\textwidth}{ |p{0.08\linewidth} | p{0.10\linewidth} | p{0.10\linewidth} | p{0.11\linewidth} | p{0.11\linewidth} | p{0.17\linewidth} |}
 \hline
Data set & Users & Groups & Roles & Data stores & Permissions \\
 \hline
 1 & 1,648 & 38 & 1,093 & 230 & 18,537 \\
2 & 180 & 30 & 673 & 167 & 7,003 \\
3 & 107 & 38 & 580 & 147 & 4,540 \\
4 & 189 & 20 & 1,117 & 452 & 16,894 \\
5 & 103 & 22 & 525 & 166 & 14,624 \\
6 & 1,587 & 51 & 794 & 257 & 30,930 \\
7 & 29 & 21 & 283 & 182 & 38,592 \\
8 & 670 & 23 & 131 & 115 & 210,645 \\
9 & 388 & 100 & 2,146 & 406 & 762,461 \\
10 & 832 & 102 & 4,902 & 1,644 & 245,860 \\
11 & 129 & 10 & 112 & 94 & 11,639 \\
12 & 674 & 17 & 1,676 & 1,654 & 1,086,714 \\
13 & 582 & 110 & 64,880 & 2,857 & 12,150,172 \\
14 & 1,113 & 132 & 15,520 & 3,477 & 1,277,689 \\
 \hline
\end{tabular*}
\end{center}
\caption{Properties of real-world data sets}
\label{table:data_sets}
\end{table}

\section{PDDL Definition for Actions}

We next list PDDL definitions of some of the actions introduced in the main text.

Listing~\ref{lst:permFlow_4tpl} shows how permissions flow for 4-tuple relations. The interpretation is analogous to action \texttt{permissionFlow\_id\_3tpl}. If $\mr{isFlowActive(\text{\id2, \id1})}$ holds, and \id2 has tuple $\mr{id\_4tpl(\text{\id2, \id3, \perm, \id4})}$, then $\mr{id\_4tpl(\text{\id1, \id3, \perm, \id4})}$ holds. Action $\mr{permissionFlow\_ds\_4tpl}$ is defined analogously for datastores.

\begin{lstlisting}[
  float=t,
  caption={\small Permission flow of 4-tuple relations from \id2 to \id1},
  label={lst:permFlow_4tpl},
  language=PDDL,numbers=none,basicstyle=\ttfamily\small]
(:action permissionFlow_id_4tpl
    :parameters (?id1 ?id2 ?id3 ?perm ?id4)
    :precondition 
    (
        id_4tpl(?id2 ?id3 ?perm ?id4) &
        isFlowActive(?id2 ?id1)&
        not id_4tpl(?id1 ?id3 ?perm ?id4)
    )
    :effect id_4tpl(?id1 ?id3 ?perm ?id4)
)
\end{lstlisting}

\paragraph{Actions for handling regular expressions} For succinct and efficient representation in PDDL (w.r.t.$\!$ the number of tuples), we use the keyword ``full\_control" to match to \textit{any} permission in the system, and ``any\_user" to match to any user (or analogously any role/group) in the cloud environment. Such permissions are useful for defining admin users or admin policy, or succinctly defining a set of permissions an identity can have. This representation also helps the planner as the planner only needs to instantiate the required tuple for an attack to match these keywords (and not explicitly create all the tuples). It is shown in supplement.

Action in Listing~\ref{lst:activate_id_3tpl} enables such a regular expression matching. The action checks if predicate $\mr{id\_tpl}\text{(\id1, \perm, \id2)}$ can be set to true. This is only possible if \textit{full\_control} keyword is present to match to  \perm, and any\_user matches to \id2, among other preconditions set in this action. Analogously, we can define such an action for datastores as $\mr{activate\_ds\_3tpl}$ where we can use keyword ``any\_datastore" instead of any\_user.

\begin{lstlisting}[mathescape=true,
  float=t,
  caption={\small Addressing regular expression $\star$ for ``full\_control" and ``any\_user"},
  label={lst:activate_id_3tpl},
  language=PDDL,numbers=none, basicstyle=\ttfamily\small]
(:action activate_id_3tpl
    :parameters (?id1 ?perm ?id2)
    :precondition 
    (
       not id_tpl(?id1 ?perm ?id2) &
       (
        id_tpl(?id1 full_control ?id2)|
        id_tpl(?id1 full_control any_user)|
        id_tpl(?id1 ?perm any_user)
       )
    )
    :effect id_tpl(?id1 ?perm ?id2)
)
\end{lstlisting}

\paragraph{Actions to enable attacks}
Listing~\ref{lst:attacks} shows \texttt{moveObject} action for the sensitive data flow attack. This action moves data from datastore \ds1 to \ds2 by copying data from \ds1, moving it to \ds2, and then deleting data from \ds1. We also check if \ds1 has sensitive data as only this enables sensitive data exfiltration attack. In the action effect, if \ds2 is a public datastore, then it makes the attack predicate true. 

\begin{lstlisting}[
  float=t,
  caption={\small Sensitive data attack action},
  label={lst:attacks},
language=PDDL,numbers=none,basicstyle=\ttfamily\small]
(:action moveObject
    :parameters (?id ?ds1 ?ds2 )
    :precondition (compromised_id ?id)  &
        has_sensitive_data(?ds1)        &
        (not is_dummy_datastore(?ds1))  &
        (not is_dummy_datastore(?ds2))  &
        ds_tpl(?id s3_GetObject ?ds1)   &
        ds_tpl(?id s3_DeleteObject ?ds1)& 
        ds_tpl(?id s3_PutObject ?ds_2)
    :effect has_sensitive_data(?ds2)    &
        (not has_sensitive_data(?ds1))  & 
        (when (is_public_datastore ?ds_2)  
              (sensitive_data_exfiltration)
        )
)
\end{lstlisting}

\paragraph{Actions to select compromised user}
We also need to define which user is compromised. There can be multiple ways to do this. We can run the planner by setting each user as compromised one by one. Another method, which is faster, when we are looking to generate any attack plan for a particular type of attack is to let the planner choose which user to make compromised. This action is \texttt{selectCompromisedUser} shown in Listing~\ref{lst:selectCompromised}. Typically, we compromise user identities (not group or role-type identities). We also have additional predicates to make sure that the action \texttt{selectCompromisedUser} is the first action to be executed (and only once) in the plan (details omitted). It is also possible to compromise multiple users in case required by the end use case.

\begin{lstlisting}[
  float=tb,
  caption={\small Action to select a compromised user},
  label={lst:selectCompromised},
  language=PDDL,numbers=none,basicstyle=\ttfamily\small]
(:action selectCompromisedUser
    :parameters ?id
    :precondition user_pred(?id)
    :effect compromised_id(?id)
)
\end{lstlisting}

\section{NP-Hardness: Reducing Set Cover to Attack Path}

Consider the set cover problem which is NP-Hard. We are given a set $V=\{v_1, \ldots, v_n\}$ of $n$ elements. We are also given $m$ subsets $S_1$ to $S_m$ of $V$. Goal is to find the minimum number of subsets whose union covers all elements in $V$. We reduce this problem to our IAM attack path problem.

The set of identities is $\{S\} \cup \{S_1, \ldots, S_m\}$. We create an identity for each set $S_i$, and an additional identity $S$.

We have $n$ datastores $\{v_1, \ldots, v_n\}$.

There are two permissions $\text{\perm} \in \{\mathrm{canAssume, hasElement}\}$. The intuition is that when we have tuples $\<S, hasElement, v_i\> \forall i=1:n$, then it means that we have found a cover. And the cost of each action is 1, so only minimum number of subsets $S_j$ will be used in the optimal attack plan.

The initial state $s_0$ is a set of below predicates:
\begin{itemize}
    \item $\compromised(S)$. That is, only $S$ is compromised id.
    \item $\<S, canAssume, S_j\> \; \forall j=1:m$
    \item $\<S_j, hasElement, v_i\> \; \forall v_i\in S_j, \forall S_j$
\end{itemize}

The preconditions are described below:
\begin{itemize}
    \item $\mr{isFlowActive}(S_j, S;s)$ requires $\<S, \mr{canAssume}, S_j\>\in s$
    \item There is only one attack type called $\mr{setcover}$. Its precondition is: 
    \begin{align}
    \mr{precondition}(\text{\id}, \mr{setcover}; s)=\nonumber \\
    \Large\bigwedge_{i=1}^n \{\<S, hasElement, v_i\>\} \bigwedge \{\compromised(\text{id})\} \in s
    \end{align}
\end{itemize}

In this formulation, the only way for the id $S$ to get tuples $$\{\<S, hasElement, v_i\>\}$$ is via the id $S$ assuming the role of some subset $S_j$. Since subsets $S_j$ have permissions $\<S_j, hasElement, v_i\>$, once $S$ assumes the role $S_j$, $S$ will get tuples $\<S, hasElement, v_i>$. As each action has cost 1, the planner will only use minimum number of assumeRole operations. Only those $S_j$ that are part of an assumeRole operation, will be part of optimal set cover. Furthermore, since precondition for the attack is id $S$ must have all tuples $\<S, hasElement, v_i\> \forall i=1:n$, it means that all elements of the set must be covered.

\begin{table*}[tbp]
    \centering
    \begin{tabular}{|p{0.07\textwidth}|p{0.08\textwidth}|p{0.20\textwidth}|p{0.21\textwidth}|p{0.3\textwidth}|}
        \hline
         Service & Resource Type & AWS API & PDDL Action/Predicate & Description\\
         \hline
         IAM & User & CreateUser & gainPersistenceAction & Create an IAM user \\
         \cline{3-5}
         &  & CreateLoginProfile & changeUserLogin & Create a password for an IAM user \\
         \cline{3-5}
         &  & UpdateLoginProfile & gainPersistenceAction & Update the password for an IAM user \\
         \cline{3-5}
         &  & PutUserPolicy & identity\_4tuple\_pred & Attach an inline policy to a user \\
         \cline{3-5}
         &  & DeleteUserPolicy & DeleteUserPolicy & Remove an inline policy from a user \\
         \cline{3-5}
         &  & AttachUserPolicy & identity\_4tuple\_pred & Attach a managed policy to a user \\
         \cline{3-5}
         &  & DetachUserPolicy & DetachUserPolicy & Detach a managed policy from a user \\
         \cline{3-5}
         &  & ChangePassword & changeUserLogin & Change the IAM user's password \\
         \cline{3-5}
         &  & CreateAccessKey & identity\_4tuple\_pred & Create a new access key for a user \\
         \cline{3-5}
         &  & DeleteAccessKey & DeleteAccessKey & Delete an access key from a user \\   
         \cline{3-5}
         &  & UpdateAccessKey & UpdateAccessKey & Modify the status of an access key \\
         \cline{3-5}
         &  & DeactivateMFADevice & DeactivateMFADevice & Deactivate an MFA device for a user \\
         \cline{2-5}
         
         & Group & DeleteGroup & DeleteGroup & Remove an IAM group \\
         \cline{3-5}
         &  & PutGroupPolicy & identity\_4tuple\_pred & Attach an inline policy to a group \\
         \cline{3-5}
         &  & AttachGroupPolicy & identity\_4tuple\_pred & Attach a managed policy to a group \\
         \cline{3-5}
         &  & AddUserToGroup & identity\_4tuple\_pred & Add a user to a specific group \\
         \cline{3-5}
         &  & RemoveUserFromGroup & RemoveUserFromGroup & Remove a user from a specific group \\
         \cline{2-5}
         
         & Role & AssumeRole & permissionFlow* & Assume an IAM role to access resources \\
         \cline{3-5}
         &  & UpdateAssumeRolePolicy & identity\_4tuple\_pred & Modify the trust policy for a role \\
         \cline{3-5}
         &  & DeleteRole & DeleteRole & Remove an IAM role \\
         \cline{3-5}
         &  & PutRolePolicy & identity\_4tuple\_pred & Attach an inline policy to a role \\
         \cline{3-5}
         &  & DeleteRolePolicy & DeleteRolePolicy & Remove an inline policy from a role \\
         \cline{3-5}
         &  & AttachRolePolicy & identity\_4tuple\_pred & Attach a managed policy to a role \\
         \cline{3-5}
         &  & DetachRolePolicy & DetachRolePolicy & Detach a managed policy from a role \\
         \cline{2-5}
         
         & Policy &  DeletePolicy &  DeletePolicy & Remove a managed IAM policy \\
         \cline{3-5}
         &  & CreatePolicyVersion &  identity\_4tuple\_pred & Create a new version of a policy \\         
         \cline{2-5}
         
         \hline
         Lambda & Function & CreateFunction & gainPersistenceAction & Create a new Lambda function \\
         \cline{3-5}
         &  & UpdateFunctionCode & gainPersistenceAction & Update the code of a Lambda function \\
         \hline
         EC2 & Instance & RunInstances & gainPersistenceAction & Launch new EC2 instances \\
         \cline{3-5}
         &  & ModifyInstanceAttribute & gainPersistenceAction & Change an attribute of an EC2 instance \\
         \hline
         
         S3 & Bucket & GetObject & copyObject, moveObject & Retrieve an object from an S3 bucket \\
         \cline{3-5}
         &  & PutObject & copyObject, moveObject & Upload an object to an S3 bucket \\
         \cline{3-5}
         &  & DeleteObject & moveObject & Remove an object from an S3 bucket \\
         \cline{3-5}
         &  & CreateBucket & createPublicBucket & Create a new S3 bucket \\
         \cline{3-5}
         &  & DeleteBucket & DeleteBucket & Remove an S3 bucket \\
         \cline{3-5}
         &  & PutBucketAcl & createPublicBucket & Set the ACL for an S3 bucket \\
         \hline
         SSM & Command & SendCommand & gainPersistenceAction & Run commands on managed instances \\
         \cline{3-5}
         &  & StartSession & gainPersistenceAction & Start a session for remote management \\
         \hline
         KMS & Key & CreateKey & encryptSensitiveData & Create a new KMS key \\
         \hline                  
    \end{tabular}
    \caption{Modeled AWS APIs and their respective description.}
    \label{table:aws_apis_desc}
\end{table*}

\section{Modeled AWS APIs and their semantics}
AWS APIs modeled by \sysname are listed in Table~\ref{table:aws_apis_desc}.
{We also list the corresponding PDDL action or predicate that each AWS API is associated with (3rd column).
Most AWS APIs are associated with a single PDDL action, while some are used to construct 4-ary predicates, 
such as identity\_4tuple\_pred. The AssumeRole API is used in all permissionFlow predicates.
}


